\documentclass{PoS}

\def\LP{\left(}		% left parenthesis
\def\RP{\right)}	% right parenthesis
\def\BE{\begin{displaymath}}
\def\EE{\end{displaymath}}
\def\BEA{\begin{eqnarray*}}
\def\EEA{\end{eqnarray*}}
\def\BNEA{\begin{eqnarray}}
\def\ENEA{\end{eqnarray}}

\newcommand{\Dslash}{\makebox[0pt][l]{\,/}D}

\title{Simulations with dynamical HISQ quarks}

\ShortTitle{Simulations with dynamical HISQ quarks}

\author{
A.~Bazavov,$^a$
\hspace{-1.0mm}\thanks{Present address: Department of Physics, Brookhaven National
Laboratory, Upton, NY 11973, USA}\,
\,C.~Bernard,$^b$
C.~DeTar,$^c$
W.~Freeman,$^a$
Steven~Gottlieb,$^d$
U.M.~Heller,$^e$
J.E.~Hetrick,$^f$
J.~Laiho,$^g$
L.~Levkova,$^c$
M.~Oktay,$^c$
J.~Osborn,$^h$
R.L.~Sugar,$^i$
\speaker{D.~Toussaint$^a$}
and
R.S.~Van~de~Water$^j$
\\
\llap{$^a$} Physics Department, University of Arizona, Tucson, AZ 85721, USA\\
\llap{$^b$} Department of Physics, Washington University, St. Louis, MO 63130, USA\\
\llap{$^c$} Physics Department, University of Utah, Salt Lake City, UT 84112, USA\\
\llap{$^d$} Department of Physics, Indiana University, Bloomington, IN 47405, USA and National Center
for Supercomputing Applications, University of Illinois, Urbana, IL 61801, USA\\
\llap{$^e$} American Physical Society, One Research Road, Ridge, NY 11961, USA\\
\llap{$^f$} Physics Department, University of the Pacific, Stockton, CA 95211, USA\\
\llap{$^g$} SUPA, School of Physics and Astronomy, University of Glasgow, Glasgow G12 8QQ, UK\\
\llap{$^h$} Argonne Leadership Computing Facility, Argonne National Laboratory, Argonne, IL 60439, USA\\
\llap{$^i$} Department of Physics, University of California, Santa Barbara, CA 93106, USA\\
\llap{$^j$} Department of Physics, Brookhaven National Laboratory, Upton, NY 11973, USA\\
E-mail: 
\email{obazavov@quark.phy.bnl.gov},
\email{cb@wuphys.wustl.edu},
\email{detar@physics.utah.edu},
\email{wfreeman@physics.arizona.edu},
\email{sg@indiana.edu},
\email{heller@ridge.aps.org},
\email{jhetrick@uop.edu},
\email{jlaiho@fnal.gov},
\email{ludmila@physics.utah.edu},
\email{oktay@physics.utah.edu},
\email{osborn@alcf.anl.gov},
\email{sugar@physics.ucsb.edu},
\email{doug@physics.arizona.edu},
\email{ruthv@bnl.gov}
}

\abstract{
We report on the status of a program of generating and using 
configurations with four flavors of dynamical quarks, using the HISQ action.
We study the lattice spacing dependence of physical quantities in
these simulations,
using runs at several lattice spacings, but with
the light quark mass held fixed at two tenths of the strange quark
mass.  We find that the lattice
artifacts in the HISQ simulations are much smaller than those in
the asqtad simulations at the same lattice spacings and quark masses.
We also discuss methods for setting the scale, or assigning a lattice
spacing to ensembles run at unphysical parameters.
          }

\FullConference{The XXVIII International Symposium on Lattice Field Theory, Lattice2010\\
		June 14-19, 2010\\
		Villasimius, Italy}

\begin{document}

\section{Introduction}

\begin{figure}
\vspace{-0.5in}
\begin{center}
\includegraphics[width=0.65\textwidth]{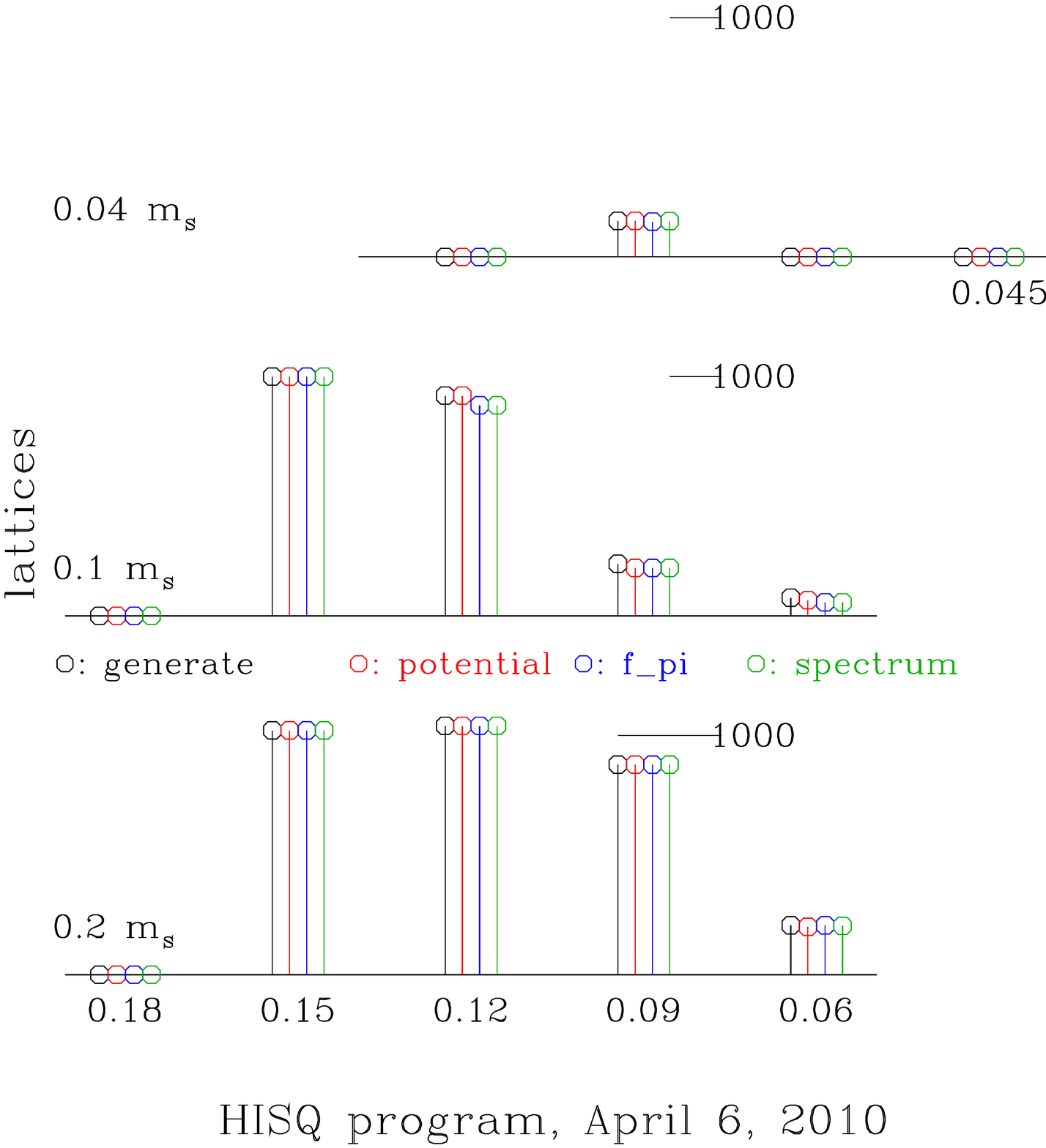}
\end{center}
\vspace{-0.4in}
\caption{
Status of the HISQ configuration generation program as of June 2010.
We expect to eventually bring all these ensembles to approximately 1000
equilibrated configurations.  The ensembles at the physical quark mass ($0.04\,m_s$)
at $a=0.06$ and $0.045$ fm will require next generation computers.
The black bars show the number of equilibrated configurations, and the red, blue
and green bars show the progress of static potential, light pseudoscalar 
amplitudes, and simple hadron spectrum calculations.  These are being done
in parallel with the configuration generation to the extent possible.
% prepare with
% axis sc 1.27 u 0.02 r .04 < status_hisq_june10.ax | plot -Tps > status_hisq_june10.ps ; ps2pdf status_hisq_june10.{ps,pdf}
\label{fig:hisq_status}}
\end{figure}

During the past eleven years the MILC collaboration has carried out a program
of QCD simulations using an improved staggered quark action, the ``asqtad''
action.  This action has lattice artifacts that are much smaller than those
in the original (one-link) staggered quark action.   These simulations used
two light and one strange dynamical quarks.  Lattices have been generated
with spacings ranging from 0.18 fm to 0.045 fm and light quark masses ranging
from 0.05 $m_s$ to $m_s$.   The spatial sizes of these lattices ranged from
2.5 fm to 5.8 fm.  In total, around 25000 configurations have been archived.  These
configurations have been used for a wide variety of QCD studies and are all publicly
available from the NERSC Gauge Connection archive, from ILDG, or informally.
A detailed review of this simulation program can be found in Ref.~\cite{RMP}.

Recently the HPQCD/UKQCD collaboration introduced a ``highly improved staggered
quark'' (HISQ) action which further reduces taste symmetry violations, which are
expected to be the largest lattice artifacts with staggered quark actions.  Using
this action, and taking advantage of improved machine power and algorithm development,
we have begun a new program of lattice QCD simulations.  
There are several differences between this new set of simulations and those in the
asqtad program.
First, the smeared links used in the $\Dslash$ operator are done
with a two level smearing, so that the quarks effectively see a smoother
lattice, leading to reduced taste symmetry violations.   Second, the
three-link (Naik) term in $\Dslash$ for the charm quark is modified to
improve the dispersion relation of the free charm quark.  These two
improvements together make up what is
usually called the HISQ action\cite{HISQACTION}.
In addition, this generation of simulations is using spatial lattice sizes
about 20\% larger than the comparable asqtad action simulations.  While this
seems like a small adjustment, since finite size effects are expected to
decrease as $\exp(-m_\pi L)$, these effects should be significantly reduced.
Also, the Symanzik improved gauge action contains the one-loop effects of
the HISQ fermions\cite{HISQGAUGE}.  The comparable asqtad effects were not available when that
program was started, but turned out to be surprisingly large\cite{ASQTADGAUGE}.
With experience from the asqtad simulations, we expect to do a better job
of tuning the dynamical quark masses, reaching a level of perhaps 2\% accuracy.
In contrast, some of the early asqtad lattice ensembles had strange quark
masses as much as 20\% off their {\it a posteriori} correct value.
Finally, we are including a dynamical charm quark.  At our smaller lattice
spacings, $a m_c$ is not large.  We do expect the effects of dynamical charm
to be small in most cases, but they may be important to the high temperature
QCD part of our program, and they are quite cheap in include.

In this program we expect to produce full scale ensembles with about 1000 
equilibrated configurations at lattice spacings
of $0.15$, $0.12$, $0.09$ and $0.06$ fm, with light quark masses of $m_s/5$,
$m_s/10$ and the physical value, roughly $m_s/27$.  In the not too distant
future we also expect to generate an ensemble with $a=0.045$ fm and the physical
quark masses.  We will also generate a number of coarser ensembles, mainly
for setting the scale in HISQ high temperature QCD simulations.
For each of these approximate lattice spacings we are using values
of $10/g^2$ and $u_0$ determined from the $m_l=m_s/5$ runs at all light
quark masses.  The strange and charm quark masses are adjusted from
short tuning runs to give the physical values of $2M_K^2-M_\pi^2$ and
$\frac{3}{4} M_\Psi + \frac{1}{4} M_{\eta_c}$ respectively.
Figure~\ref{fig:hisq_status} shows the current (June 2010) status of
these runs.

\section{Tests of scaling}

The first stage in this program was to generate complete ensembles at
a fixed but unphysical light quark mass of $m_s/5$ at lattice spacings
of $0.15$, $0.12$ and $0.09$ fm.  
This allows us to test scaling, or dependence of calculated
quantities on the lattice spacing.  In particular, we wish to see if
the reductions in taste symmetry violations are accompanied by reductions
in lattice spacing dependence of other quantities.  The results of these
tests are briefly summarized here, but a complete report may be found
in Ref.~\cite{SCALING09}.

\begin{figure}
\vspace{-0.5in}
\begin{center}
\includegraphics[width=0.65\textwidth]{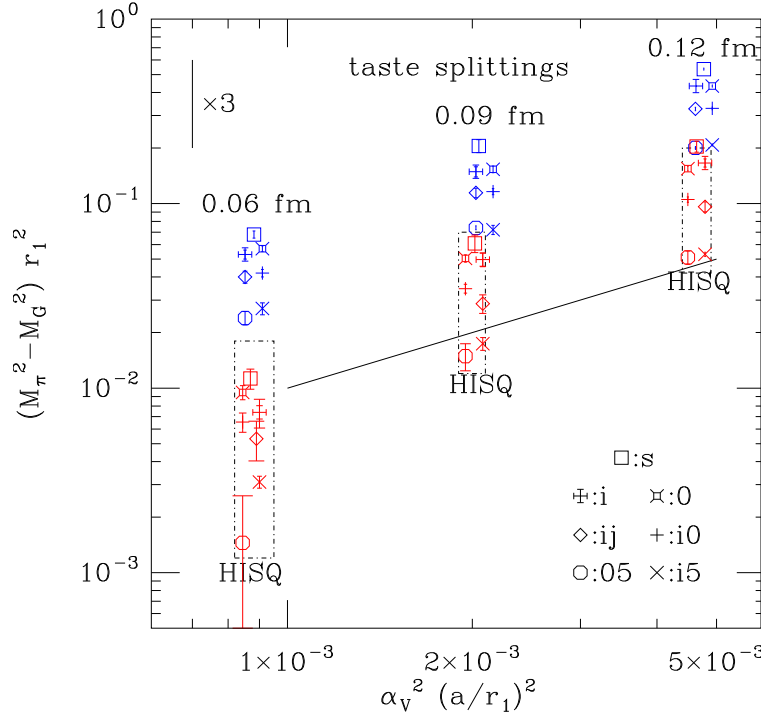}
% prepare with
% axis sc 1.10 r .08 u 0 h 1 w 1 < taste_split_try.ax | plot -Tps > taste_split_try.ps ; ps2pdf taste_split_try.{ps,pdf}
\end{center}
\vspace{-0.5in}
\caption{Taste splittings among the pions, comparing the asqtad
and HISQ actions.
%Note that taste splittings were studied earlier by the HPQCD
%collaboration\protect\cite{HISQACTION}
%using quenched and asqtad sea quarks.
The asqtad results used
configurations with 2+1 flavors of dynamical quarks, and the HISQ results
2+1+1 flavors. The
quantity plotted is $r_1^2 \LP M_\pi^2 - M_G^2 \RP$, where $M_\pi$
is the mass of the non-Goldstone pion and $M_G$ is the mass of the
Goldstone pion.
Reading from top to bottom, the non-Goldstone pions
are the $\pi_s$ (box), $\pi_0$ (fancy box), $\pi_i$ (fancy plus),
$\pi_{io}$ (plus), $\pi_{ij}$ (diamond), $\pi_{i5}$ (cross)
and $\pi_{05}$ (octagon).
$r_1^2 \LP M_\pi^2 - M_G^2 \RP$ is known to be almost independent of
the light-quark mass.
The vertical bar at the upper left shows the size of a factor of
three, roughly the observed reduction in taste splittings, while
the sloping solid line shows the theoretically expected dependence on lattice
spacing.  Nearly degenerate points have been shifted horizontally to improve
their visibility.
\label{fig:taste_split}}
\end{figure}

\begin{figure}
\vspace{-0.5in}
\begin{center}
\includegraphics[width=0.65\textwidth]{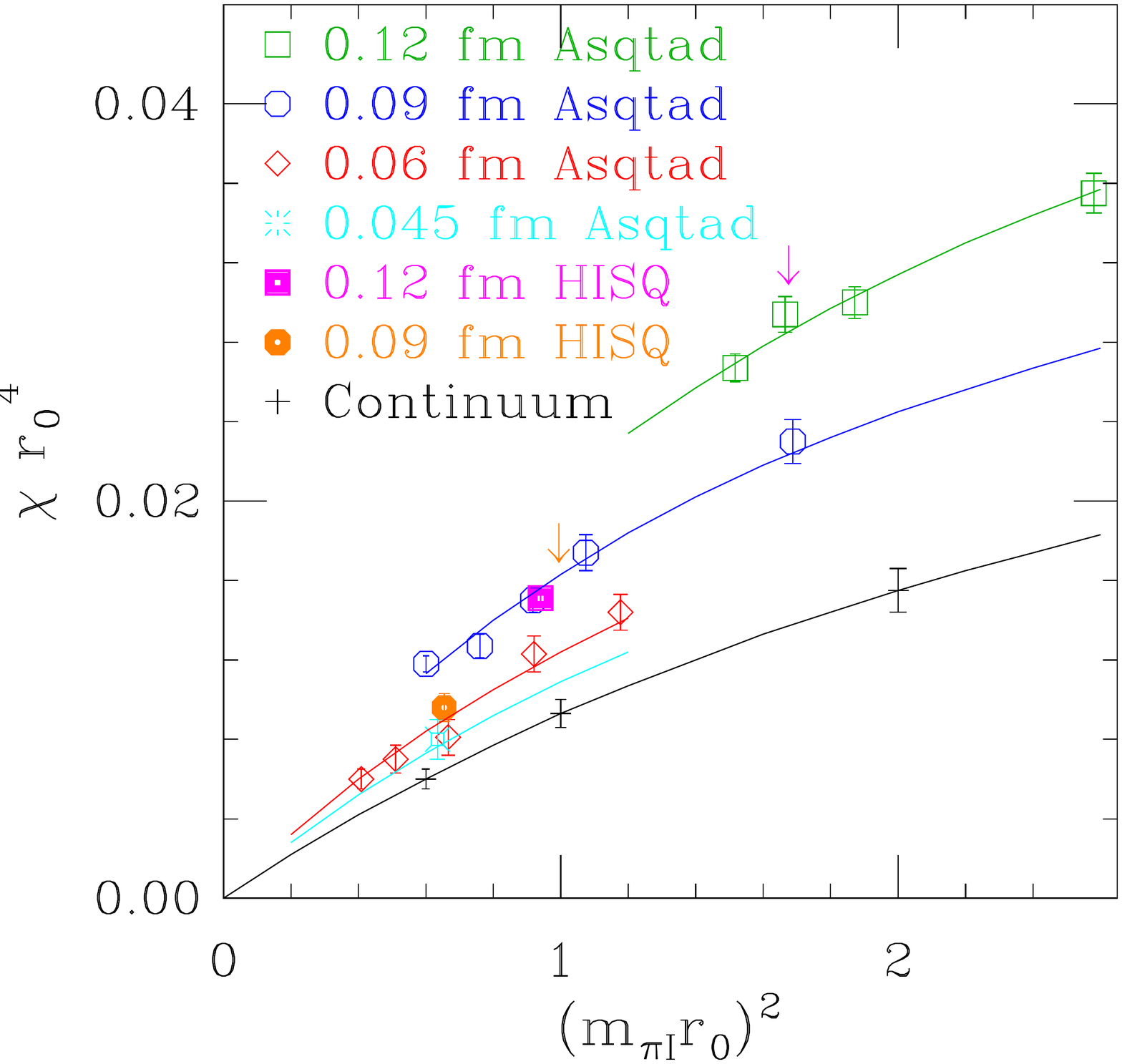}
\vspace{-0.5in}
\end{center}
% prepare with
% axis sc 1.12 u 0.02 r .11 < topology_dec09.ax | plot -Tps > topology_dec09.ps ; ps2pdf topology_dec09.{ps,pdf}
\caption{The topological susceptibility, comparing the asqtad and
HISQ actions.
Points with the
asqtad action are shown for several lattice spacings and quark
masses, and the
HISQ results for $a \approx 0.12$ fm and $a \approx 0.09$ fm with $m_l = 0.2 m_s$.
For the horizontal axis we use the mass of the taste singlet pion, since
in lowest order chiral perturbation theory the topological susceptibility
is a function of this mass~\protect\cite{topo_form}.
The curves in the
figure come from a chiral perturbation theory fit to the asqtad data.
A discussion of the methods used for computing the susceptibility and
the asqtad results can be found in Ref.~\cite{topo_milc10}.
%are updated from Ref.~\protect\cite{topo_lat07} and
%are discussed further in Refs.~\protect\cite{milc_rmp,topo_milc10}.
The two arrows indicate the locations of asqtad points with lattice spacing and
quark mass similar to the two HISQ points. (In the case of the $a\approx 0.09$ fm
HISQ point, the quark mass falls between two of the masses of the asqtad points.)
\label{fig:topology}}
\end{figure}

The HPQCD/UKQCD collaboration demonstrated the reduction of taste violations
with HISQ valence quarks using quenched and asqtad sea quarks\cite{HISQACTION}.
As expected, we find similar results with HISQ sea quarks.  
Figure~\ref{fig:taste_split} shows results for the taste splittings
with HISQ valence and sea quarks with $m_l=m_s/5$, including some results
on a partially completed ensemble with $a=0.06$ fm.

Perhaps the most stringent test of improvement is the topological
susceptibility, since the dependence of this quantity on the sea
quark mass demonstrates that the sea quarks are modifying the
gluon configurations.   Figure~\ref{fig:topology} shows the topological
susceptibility for most of the asqtad ensembles\cite{topo_milc10}, together with the
HISQ results for $m_l=m_s/5$ at $a=0.12$ and $0.09$ fm.  It can be seen
that the HISQ points fall below the corresponding asqtad points, which
are indicated by the arrows in the figure.
Note that the HISQ points are to the left of the corresponding asqtad points.
This is because the horizontal axis is the mass of the taste singlet
pion (the heaviest pion taste), and the reduction in taste symmetry
breaking moves the points to the left.  It is the movement down relative
to the asqtad points that represents an improvement in the gluon configurations.

We also looked at the masses of the $\rho$ and nucleon in these ensembles.
Strictly, these are an unphysical vector meson and baryon with both
valence and sea light quark masses at $m_s/5$.  Thus, we can check
their dependence on lattice spacing, but without a chiral extrapolation
we cannot test their masses against experiment.  With this caution, we
found that the dependence of $r_1 M_\rho$ and $r_1 M_N$ on the lattice spacing
was much smaller than for the asqtad ensembles at comparable quark masses.
More details are in Ref.~\cite{SCALING09}.
Of course, this could be interpreted either as improved scaling of hadron
masses using $r_1$ to set the scale or improved scaling of $r_1$ using hadron
masses to set the scale.

% \begin{figure}
% \includegraphics[width=0.6\textwidth]{mrho_r1_oct09.pdf}
% \caption{Rho meson masses with asqtad and HISQ actions.
% FIXX THIS
% \label{fig:rho_mass}}
% \end{figure}

%\begin{figure}
%\includegraphics[width=0.6\textwidth]{disp_eta_c.pdf}
%\caption{Dispersion relation for the $\eta_c$ meson.
%Specifically, we show the ``speed of light'', or
%$\vec p^2 = \LP\frac{2\pi}{L}\RP^2 \vec n^2$.
%Similar results have been obtained by the HPQCD collaboration using
%$a_{tad}^2$ sea quarks\protect\cite{HPQCD_CHARM_DISP}; this is with
%HISQ sea quarks.
%FIXX THIS
%\label{fig:etac_dispersion}}
%\end{figure}
%
%\begin{figure}
%\includegraphics[width=0.6\textwidth]{charmonium.pdf}
%\caption{Charmonium spectrum.  Here we just used the simplest
%pointlike operators, with a Coulomb gauge wall source and point sink.
%Note that this does not allow us to see the $\chi_{c2}$ states
%FIXX THIS
%\label{fig:charmonium_spectrum}}
%\end{figure}

% \begin{figure}
% \includegraphics[width=0.6\textwidth]{decays_p2ms_mc.pdf}
% \caption{Decay constant of a charm-light meson.  Here the light sea
% quark mass was $m_l = 0.2 m_s$.
% Eventually, find  $f_D$, $f_{D_s}$ and $f_{D_s}/f_D$.
% FIXX THIS
% \label{fig:charm_decay}}
% \end{figure}

\section{Setting the length scale}

\begin{figure}
\vspace{-0.5in}
\begin{center}
\includegraphics[width=0.65\textwidth]{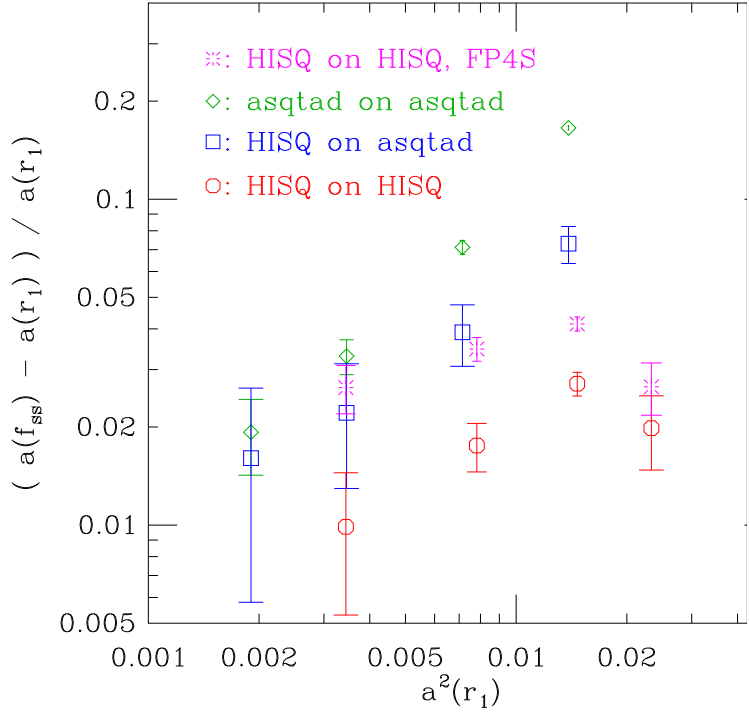}
\vspace{-0.5in}
\end{center}
% prepare with
% axis sc 1.12 u 0 r .11 < find_a_2.ax | plot -Tps > find_a_2.ps ; ps2pdf find_a_2.{ps,pdf}
\caption{
Differences in determinations of the length scale using different
standards.  In the legend, the symbol types are labelled as
``valence on sea''.
The ``FP4S'' points (bursts) use a pseudoscalar
meson with both valence quarks at 0.4 times the strange quark mass, while
all the other points use valence quarks at the strange quark mass.
The ``HISQ on asqtad'' points are taken from Ref.~\protect\cite{hpqcd_hisq_r1}.
\label{fig:find_a_2}
}
\end{figure}

QCD simulation programs involving unphysical quark masses generally involve
a definition of the lattice spacing as a function of the gauge couplings
and sea quark masses.   About the only real requirement on this definition
is that it should be correct at the physical quark masses in the continuum
limit.  The most common choice is a scale determined from the static
quark potential\cite{SOMMER}, usually either $r_0^2F(r_0)=-1.65$ or $r_1^2F(r_1)=-1$.
This is not a physical quantity, but it is a convenient interpolating quantity
that can be determined accurately with a reasonable amount of work.  Its
value is determined by matching some physical quantity, such as quarkonium
mass splittings or $f_\pi$, in the continuum and chiral limits.
Recently the HPQCD collaboration has suggested using the decay constant
of a fictitious isovector pseudoscalar meson with valence quark masses equal
to the strange quark mass as the length standard\cite{hpqcd_hisq_r1},
which we will call $f_{ss}$.
Like $r_0$ or $r_1$, this is an unphysical quantity whose value is determined
by matching a physical quantity, most likely $f_\pi$, in the continuum limit
at the physical quark mass.  Of course, one could use any mass for the
valence quark, and we have been experimenting with using a mass of $0.4$ times
the strange quark mass.  The choice of reference mass involves a tradeoff --- we
want a mass heavy enough so that an accurate simulation result is possible but
light enough so that chiral perturbation theory can be used with confidence.

Use of an unphysical decay constant has several advantages.   It is possible
to get very good accuracy, and the systematic errors from things like choices
of fit ranges are better understood than those in $r_1$.  Also, for
calculations involving hadron masses and matrix elements, there are fewer
steps in the logic relating the lattice results to dimensionful quantities.
On the other hand, the meson decay constant takes much more computer
time than the static quark potential (although it may be needed
anyway) and is dependent on the choice of valence quark 
formulations.   For example, asqtad and HISQ valence quarks would give
different lattice spacings for the same ensemble of configurations.

Table~\ref{table:fss} shows the lattice spacings for several asqtad
and HISQ ensembles determined from $r_1$ and from pseudoscalar amplitudes
with asqtad
and HISQ valence quarks.  Figure~\ref{fig:find_a_2} shows the differences
between the pseudoscalar and $r_1$ lattice spacings versus $a^2$, including
lattice spacings determined from both $f_{ss}$ and $f_{.4s}$, and it can be seen that
the differences are vanishing in the continuum limit.

\begin{table}
\begin{tabular}{|llllc|lcl|}
\hline
Action & $10/g^2$ & $am_l$ & $am_s$ & $am_c$ 	& $a(r_1)$	& $a(f_{ss,\mathrm{asq}})$	&
$a(f_{ss,\mathrm{HISQ}})$ \\
\hline
%b6572m0097m04845 & 0.1453(9)	& 0.1770(4)	& 0.1583(13) \\
% quadratic solution doesn't work for this one
asqtad & 6.76 & 0.01   & 0.05  & -- & 0.1178(2)	& 0.1373(2)	& 0.1264 (11) \\
asqtad & 7.09 & 0.0062 & 0.031 & -- & 0.0845(1)	& 0.0905(3)	& 0.0878(7) \\
asqtad & 7.46 & 0.0036 & 0.018 & -- & 0.0588(2)	& 0.0607(1)	& 0.0601(5) \\
asqtad & 7.81 & 0.0028 & 0.014 & -- & 0.0436(2)	& 0.0444(1)	& 0.0443(4) \\
\hline
HISQ & 5.80 & 0.013  & 0.065  & 0.838 & 0.1527(7)	& na	& 0.1558(3) \\
HISQ & 6.00 & 0.0102 & 0.0509 & 0.635 & 0.1211(2)	& na	& 0.1244(2) \\
HISQ & 6.30 & 0.0074 & 0.037  & 0.440 & 0.0884(2)	& na	& 0.0900(1) \\
%b672m0048m024m286	& 0.0588(4?)	& na	& 0.0602(1) \\
	\hline
	\end{tabular} \\
\caption{
\label{table:fss}
Lattice spacing determinations by various methods.  The values for
HISQ valence quarks with asqtad sea quarks are taken from
Ref.~\protect\cite{hpqcd_hisq_r1}.  In the other cases, the errors are
statistical only; they do not include the errors on the physical
value of $r_1$ ($r_1=0.3108(15)(\null_{-79}^{+26})$) or on the
estimate of $f_{ss}$ in MeV, $f_{ss}= 181.5(1.0)$ MeV\protect\cite{hpqcd_hisq_r1}.
}
\end{table}

\acknowledgments

This work was supported by the U.S. Department of Energy and National
Science Foundation.
Computation for this work was done at
the Texas Advanced Computing Center (TACC),
the National Center for Supercomputing Resources (NCSA),
the National Institute for Computational Sciences (NICS),
the National Center for Atmospheric Research (UCAR), % Frost
the USQCD facilities at Fermilab,
and the National Energy Resources Supercomputing Center (NERSC),
under grants from the NSF and DOE.
We thank Christine Davies, Alan Gray, Eduardo Follana, and Ron Horgan
for discussions and help in developing and verifying our codes.

\end{document}